# Change from within? The strategies used by public officials to advance post-growth approaches

Laura Angresius[a*], Milena Büchs[b], Daniel W. O'Neill[a, b]

[a] UB School of Economics, Universitat de Barcelona, C/ de John Maynard Keynes 1-11, 08034 Barcelona, Spain

[b] Sustainability Research Institute, School of Earth, Environment and Sustainability, University of Leeds, Leeds, LS2 9JT, United Kingdom

## Abstract

Current societies face interconnected environmental and social crises. Post-growth research argues that addressing these challenges requires a reorganization of society around the priorities of environmental sustainability, social equity, and human wellbeing over economic growth. While scholars highlight the state's potential role in enabling post-growth transformations through changes from within government institutions, post-growth-minded public officials face tensions between aspiring for radical changes of established structures while working within these structures. To understand how public officials across contexts navigate this tension, we ask: How do post-growth-minded public officials promote post-growth approaches in their work? How do the strategies differ between civil servants and elected officials? What do these strategies reveal about the capacity of the state to advance post-growth transformations? To answer these questions, we interviewed 41 post-growth-minded civil servants and elected officials. Interviews covered seven European countries and local to supranational governance scales.

We find that public officials aim to influence thinking and discourses as well as decision-making processes and the implementation of policies. Overall, elected officials tend to feel that they can be more outspoken in their activities whereas civil servants are more inclined to promote post-growth approaches indirectly. Both groups pursue coalitions with various actors within and beyond their institutions. Public officials' strategies underscore the limited capacity of the state to advance post-growth approaches under the current growth paradigm. We suggest that cooperation with civil society actors is central to build a sense of collective agency and to foster the interactions between symbiotic and interstitial strategies for post-growth transformations.

## 1. Introduction

Current societies are confronted with multiple environmental and social crises (Fanning & Raworth, 2025; Sakschewski et al., 2025). Post-growth research suggests that in order to overcome these crises there is a need to reorganise societies around environmental sustainability, social equity, and human wellbeing, instead of the continuous pursuit of economic growth (Hickel, 2020; Hofferberth, 2025; Jackson, 2017). We use *post-growth* as an umbrella term for a variety of growth-critical visions of societal transformation, including degrowth, the wellbeing economy, and Doughnut economics (Gerber & Raina, 2018; Kallis et al., 2025; Van Eynde et al., 2026). An open question remains around how post-growth transformations can be facilitated in practice, and specifically which role the state can play in this process (Barlow et al., 2022; Hofferberth, 2025).

In academic debates on post-growth transformations the state holds an ambivalent role: it has been identified as a transformative actor for societal changes (D'Alisa & Kallis, 2020; Koch, 2020b) as well as part of the problem (Brand et al., 2025; Fitzpatrick, 2024; Keyßer et al., 2025). The different perspectives either foreground the state's regulatory power and potential to implement changes on the societal scale (Bärnthaler, 2024; Eckersley, 2021) or argue that the state is structurally incapable of overcoming the status quo (Blühdorn, 2020; Hausknost, 2020). Anarchist perspectives even argue that the exercise of state power itself is running contrary to the values of a degrowth society (Fitzpatrick, 2024).

Erik Olin Wright's logics of social transformation is a prominent framework in the degrowth literature that addresses how government institutions could contribute to societal changes (Chertkovskaya, 2022; Wright, 2010, 2019). While the framework does not provide a theory of change for post-capitalist transformations (Stevenson, 2025), it distinguishes different types of strategies and describes their interactions. Wright differentiates between ruptural, interstitial, and symbiotic strategies to transform the capitalist system: Ruptural strategies seek to overcome capitalism through revolutionary tactics. Interstitial approaches aim to build alternative spaces and practices that are organized in a non-capitalist logic. Symbiotic approaches work within established institutions and political structures to transform the current system from within. For interstitial and symbiotic strategies to mutually support each other, political actors in established institutions and civil society actors have to cooperate and compromise (Harris & Jervis, 2025; Wright, 2010, 2019).

There are a range of post-growth-leaning government initiatives (henceforth just *post-growth initiatives*) that attempt to foster policy changes (Hough-Stewart & Janoo, 2024; Schmid, 2025). For example, in Wales, which is one of the Wellbeing Economy Governments, the Future Generations Act was adopted in 2015, obliging public bodies to take their decisions in line with future generations' wellbeing (Messham & Sheard, 2020; Welsh Government, n.d.). Moreover, various local governments across the world have adopted Doughnut economics frameworks to foster internal, procedural changes (Grcheva

& Vianello, 2025). However, despite efforts by post-growth proponents to influence policymaking (European Parliament, 2023; Kallis et al., 2023) so far post-growth ideas have had limited impact within governments (Kranke, 2025).

Empirical research suggests post-growth initiatives face significant structural barriers, including the dominance of the economic growth paradigm, siloed ways of working, and a lack of capacity and resources (Angresius et al., 2025; Hayden, 2025; Mason & Büchs, 2023; Schmid, 2025). Given their position in the state, public officials can address these barriers with levers that are inaccessible to outside actors such as control over budgets and formal decision-making powers (Abers, 2021). Furthermore, policy is inherently shaped by the individuals practically working on its formulation, adoption, and implementation within political institutions (Hilbrandt, 2025). However, public officials aiming to address these barriers face a tension between aiming for radical changes of government institutions while working in and for them (Porpora, 2016; Vestergaard & Schmid, 2026).

So far, the post-growth literature has not specifically examined strategies of public officials in fostering post-growth approaches and in addressing the tension of aiming for radical changes from within institutions that represent the status quo. Existing empirical work on ways in which public officials engage with post-growth approaches either examines specific policy fields (Lamker & Dieckhoff, 2022) or specific local cases (Vestergaard & Schmid, 2026). Studies on public officials' strategies across contexts are lacking. To better understand the potential role of public officials and the state in promoting post-growth approaches, we ask the following three questions: (1) How do post-growth-minded public officials promote post-growth approaches in their work? (2) How do strategies differ between civil servants and elected officials? (3) What do these strategies reveal about the capacity of the state to advance post-growth transformations?

In this study, we use *strategies* to refer to the concrete, purposeful actions public officials employ in their everyday work to foster post-growth approaches. We use *public officials* as an overarching term that includes elected officials and civil servants. By *elected officials* we mean members of the executive and/or legislative branches of government who have been elected by the public. *Civil servants* refer to non-elected members of staff of the executive branches of government. While the exact understandings of their roles in the literature depend on the theoretical approach taken, civil servants tend to be seen as non-partisan and responsible for technical implementation of policies. In contrast, elected officials take party political positions and thus openly support specific ideologies and party interests (Hindkjaer Madsen, 2024; Svara, 2001, 2006; Weber, 2009).

Strengthening post-growth approaches from within political institutions involves civil servants as well as elected officials. Based on their different institutional roles, the strategies between the two groups might differ. Our study therefore compares the strategies of civil servants and elected officials.

This article proceeds as follows: Section 2 summarizes the literature on the role of the state in post-growth transformations, presents empirical findings about government-led post-growth initiatives, and change agents' strategies in political institutions. Section 3 outlines the study's methods, and Section 4 presents the findings. Section 5 discusses our contribution to the post-growth literature and learnings from the findings. Section 6 concludes.

## 2. The roles of the state and public officials for post-growth transformations

In the following section we summarize the existing literature on the role of the state in promoting post-growth transformations and outline the tension that post-growth-minded public officials face when aiming to change established institutional structures and policy priorities from within governmental institutions. We identify an important gap in the post-growth literature on the role of the state related to the strategies of civil servants and elected officials to foster post-growth approaches across Europe.

### 2.1 The role of the state for post-growth transformations

The post-growth literature identifies various possible trajectories of post-growth transformations and respective strategies to foster them (Barlow et al., 2022; Brand & Wissen, 2017; Buch-Hansen & Nesterova, 2023). Within the post-growth movement, the plurality of strategies and tactics considered ranges from anarchist organizing to regulatory reformism (Fitzpatrick et al., 2025; Kommandeur et al., 2025). The understanding of the role of the state in post-growth transformations differs depending on the theoretical and ideological stance taken (Brand, 2016; Fitzpatrick et al., 2025). For instance, anarchist positions in the degrowth movement understand the dismantling of state power as a precondition for, and a characteristic of, degrowth societies (e.g. Fitzpatrick, 2024). Moreover, some authors argue that the transformative potential of state action for sustainability transformations is limited by state imperatives within capitalism forcing government institutions to uphold the current economic and political system including continuous economic growth (e.g. Blühdorn, 2020; Brand et al., 2025; Hausknost, 2020).

However, some post-growth scholars from a materialist state theory tradition see the state as an important field of post-growth transformations (relating to struggles within existing government institutions), and as essential to bringing about social-ecological changes on the societal scale (through respective regulations) (Bärnthaler, 2024; D'Alisa & Kallis, 2020; Koch, 2020b). They understand civil society and state actions as closely linked and both crucial to fostering societal change (Koch, 2022). Frictions and conflicts between different portfolios in, and branches of, the state could be entry points for changes in institutional structures (Koch, 2020b; Wright, 2010). Thus, from this perspective there is a potential for government institutions to change their structures and functions away from a complicity with the growth paradigm towards post-growth-aligned functions (Hasselbalch & Kranke, 2023; Koch, 2020b).

Empirical research on existing post-growth government initiatives suggests that initiatives, such as Wellbeing Economy Governments and Doughnut economics initiatives, have rather limited transformative impact (Hayden, 2025; Mason & Büchs, 2023). The limitations of the state in promoting post-growth approaches include existing growth-dominant discourses, norms and practices, as well as practical barriers like siloed ways of working and short-term thinking in policymaking. Empirical research also points to a number of enablers for post-growth government initiatives, such as coalition building and participatory governance (Angresius et al., 2025; Bärnthaler et al., 2024; Schmid, 2025).

Two further enablers for post-growth government initiatives are the actions of individual public officials within political institutions and high-level political support by incumbent governments (Angresius et al., 2025; Trebeck, 2024; Turrini et al., 2026). Thus, public officials play a key role in strategizing for post-growth transformations from within government institutions. On the one hand, all individuals in government must work with and relate to structural barriers. On the other hand, public officials can also harness and seek to strengthen enablers of change, for example through the implementation of legislative frameworks prioritising wellbeing and environmental objectives (Bourdin & Jacquet, 2025; Mason & Büchs, 2023; Schmid, 2025). In the next section, we explore empirical findings on public officials' strategies to foster policy changes and the tensions they face in their institutional roles.

## 2.2 Agency and change agents within government institutions

A key question in debates about societal change revolves around the relationship between structure and agency. Structuralists argue that social structures determine individuals' behavior (e.g. Althusser, 2020) while actor and social action perspectives highlight the potential of agents to enact change (e.g. Weber, 1949). From a critical realist perspective, there is a dialectic relationship between structure and agency (Nesterova, 2024; Porpora, 2016). Agents transform or reproduce structures, while structures constrain and at times empower agents (Bhaskar, 1998; Buch-Hansen & Nielsen, 2020). While structural conditions provide the context for certain policies to emerge (or not), agent-centered explanations can investigate *how* certain policies emerged and changes took place. The actual impetus for changes must come from people (Buch-Hansen & Nielsen, 2020). Wright (2019, p. 124) argues that *"Human emancipation, if it is to come about, requires strategy, and this implies agency. And since some of the targets of such strategy are powerful institutions, an effective strategy requires collective agency"*. Thus, in Wright's understanding, only collective actors exercising collective agency are able to reshape government institutions and foster societal change.

Transition studies and public policy literature discuss the role of individuals within government institutions in fostering institutional and policy changes in various policy fields (Braams et al., 2024; Pettinicchio, 2012). We broadly frame this literature as the "change agents" literature. Different concepts foreground different aspects to explain how actors working within government institutions aim for procedural and policy changes in their institution (Braams et al., 2024). For example, *policy entrepreneurship* focuses on the search for innovative solutions and windows of opportunity (e.g. Frisch Aviram et al., 2020), the literature on *street-level bureaucrats* underscores the power of discursive interaction (e.g. Lipsky, 2010), research on *administrative guerillas* inquires subversive actions within governments (e.g. O'Leary, 2019) and *institutional activism* (e.g. Pettinicchio, 2012), and *inside activism* (e.g. Hysing & Olsson, 2018) aim to understand the transmission of social movement demands into government action.

The empirical change agents literature has identified a range of strategies that individual public officials employ. These include influencing the public agenda and garnering public support, using information and framing strategically to influence the policy process, forging inter- and intra-organizational partnerships and generating legitimacy through professionalism and moral judgement (Abers, 2021; Braams et al., 2024; Frisch Aviram et al., 2020; Hysing & Olsson, 2018). These studies include various institutions and policy fields, for example greening central banking (DiLeo et al., 2025; Siderius, 2023), advocating for women's health and environmental policy (Abers, 2021) and sustainable development (Hysing & Olsson, 2018), as well as European and non-European contexts.

However, to date there is little research on public officials' strategies to foster post-growth approaches. Vestergaard and Schmid (2026) examined institutional activism within Amsterdam's city administration through an ethnographic case study of their Doughnut economics initiative and a climate emergency letter written by civil servants. Their study highlights different strategies for institutional change, including implementing official projects and acting beyond formal mandates based on personal values. While demonstrating that transformative impulses can emerge from within state institutions, the study also reveals the tensions public officials face when challenging institutional logics from within government.

Besides the "external barriers" for post-growth government initiatives discussed above, Lamker and Diekhoff (2022) qualitatively investigate the "internal barriers" of planners in Germany trying to move away from a growth focus in their work. They underscore the relevance of emotions and the values of individuals working in public administrations for transitioning to post-growth-aligned institutional practices. They conclude that planners' self-understanding as neutral actors hinders them from engaging in politicizing, controversial debates.

Altogether, the literature on change agents suggests that public officials hold relevant levers for post-growth-aligned institutional and policy changes. However, so far, the literature on the role of the state in post-growth transformations engages little empirically with the practicalities of fostering post-growth approaches from within government institutions. While the change agents literature has investigated civil servants' strategies in various policy fields and contexts, empirical work investigating public officials' strategies to foster post-growth approaches either examines a specific local case (Vestergaard & Schmid, 2026) or a specific policy field (Lamker & Dieckhoff, 2022). In addition, despite their different institutional roles, no study has focused on the difference in strategies of civil servants and of elected officials.

In this study, we aim to address this gap in the literature by examining the strategies of civil servants and elected officials to foster post-growth approaches in government institutions across European contexts. In doing so, this article contributes to the discussion on the role of the state in promoting post-growth approaches. We ask the following three research questions: (1) How do post-growth minded public officials promote post-growth approaches in their work? (2) How do strategies differ between civil servants and elected officials? (3) What do these strategies reveal about the capacity of the state to advance post-growth transformations? The next section outlines the methods we employed to address these questions.

## 3. Methods

Qualitative research allows us to investigate process- and change-oriented questions in depth (Derrington, 2019). We chose expert interviews to gain access to first-hand descriptions of public officials' strategies (Meuser & Nagel, 2009). We conducted semi-structured interviews with civil servants and elected officials across European, national, regional, and local governance scales in seven European countries. Table A1 in the appendix provides an overview of the interviews.

### 3.1 Data collection

For this research, we sampled public officials who were working on post-growth-leaning projects (in a broad sense) and who self-identify as post-growth-minded (see details below). We aimed to include people currently or previously working as civil servants or elected officials from different contexts to be able to identify cross-cutting strategies. The interviews cover participants from seven countries across Western, Central, North, South, and Eastern Europe working at the local, regional, national, or European governance scales. The sample included people with different levels of seniority to capture actions of individuals with different levels of agency. Altogether, we conducted 54 semi-structured online interviews between November 2024 and March 2025 of which 41 were included in the final analysis.

We selected research participants in two steps. First, we reached out to personal contacts and contacts from our previous research on post-growth government initiatives (Angresius et al., 2025) with an email reading: *"We would like to talk to post-growth minded practitioners who work on wellbeing economy related projects (broadly defined) to understand which strategies they apply to foster post-growth aligned approaches in their work […]. If you feel like the description applies to you, would you be open to have a conversation?".* We extended our sample by snowballing, following suggestions from people we had invited to take part in this research. In a second step, we verified interviewees' self-classification as “post-growth minded”. To do so, we assessed their response to at least one of the following two interview questions: “(1) How would you define the term ‘post-growth economy’?[1] (2) Which role, if any, should economic growth have as a goal in policymaking or implementation?” To comply with our understanding of “post-growth-minded” the individuals needed to express a position that was critical towards the prioritization of and reliance on economic growth and/or GDP in policymaking. In addition, they had to express an ambition to prioritize wellbeing and safeguarding environmental limits over economic growth. If there were doubts about the position of the interviewee from the response to the two specific questions, the broader context of the interview was considered. In the second step of sampling, we excluded eight interviews. In addition, five context interviews with civil society actors were excluded from the final analysis.

Thus, the final sample for this study consists of 41 interviews: 24 civil servants and 17 elected officials. Ten interviewees were working in Austria, five in Croatia, six in Finland, three in the Netherlands, six in Spain, ten in the UK, and one in France.

## 3.2 Data analysis

All interviews were transcribed manually by a transcription company and then imported into the qualitative analysis software Nvivo15. After familiarising ourselves with the transcripts, the interviews were coded deductively-inductively (Braun & Clarke, 2022). We began coding with a codebook based on topics from the interview guide and literature on post-growth transformation and change agents. For example, we had asked explicitly which strategies the public officials employed, who they sought collaborations with, which terminology they used, and whether they utilised specific decision-making tools or indicator frameworks. The interview guide is available in the supplementary materials. We then used inductive coding to capture themes we had not anticipated prior to the analysis. Examples for inductive codes are the kind of information interviewees perceive as most legitimate and which audience they primarily address. The coding was an iterative process, with codes refined and merged as the analysis progressed. Where appropriate, more than one code was assigned to the same unit of meaning. The unit of meaning varied from one sentence to multiple paragraphs, depending on the context. To facilitate the comparative analysis, we wrote memos throughout the coding process to record differences and similarities between civil servants and elected officials. In addition, we used matrix coding queries in Nvivo15 to quantitatively compare the coding frequencies between civil servants and elected officials.

[1] In some interviews, we have asked for the definition of a ‘wellbeing economy’ rather than ‘post-growth economy’ since this is the framing of ‘post-growth’ used in the EU Horizon project ToBe which this research is part of.

## 4. Results

First, we present the strategies of public officials as a comparison between elected officials and civil servants (Section 4.1). Second, we describe what these strategies reveal about the capacity of the state to advance post-growth transformations (Section 4.2).

### 4.1 Strategies of public officials

Overall, we find that civil servants tend to adopt more inward-looking activities that aim to influence the discourses within their institution. They often build networks within government and cooperate with civil society actors to strengthen post-growth positions inside their institution. They tend to be cautious to openly criticize economic growth and avoid using explicit post-growth framing. In addition, the kind of information that many of them perceive as the most legitimate to support their argumentation is technical knowledge and formal agreements. Moreover, the changes that they aim to achieve are mostly tangible changes in concrete projects.

By contrast, elected officials tend to aim to influence discourses in the broader public and among other elected officials, for example in parliamentary or local council debates. They are more open to directly criticizing economic growth and using post-growth as a term. They were more likely to draw on scientific findings, best practice examples and their political program to legitimize their post-growth position. Through coalition building with social movements as well as a variety of other political actors they focus more on adopting and implementing policies and participatory governance processes that change citizens lived experiences. Table 1 summarizes the findings.

To answer our first research question on the strategies of public officials, we present our findings in two themes that were constructed inductively: (1) Strengthening support for post-growth thinking and influencing discourses (Section 4.1.1) and (2) Influencing decision-making and implementation (Section 4.1.2). Moreover, to answer our second research question on the differences between civil servants and elected officials, we organize our findings as a comparison between the two groups. While we focus on the differences, these should be understood as tendencies and not as strict binaries.

In the following we describe the findings for both themes for civil servants and for elected officials.

**Table 1:** The strategies that civil servants and elected officials employ to foster post-growth approaches in policymaking

| | Civil servants | Elected officials |
|---|---|---|
| **1. Strengthening support for post-growth thinking and influencing discourses** | | |
| **Which audiences do they address?** | Civil servants in their institution | Broader public<br><br>Other elected officials |
| **Which terminology do they use?** | More cautious of directly criticizing economic growth<br><br>Avoid using post-growth framing | More open to directly criticizing economic growth<br><br>More open to using post-growth framing |
| **What kind of information do they perceive as most legitimate?** | Technical evidence from governance tools, indicator frameworks and academic advice<br><br>Policies, agreements, and legislation | Scientific findings<br><br>Best practice examples<br><br>Political program |
| **2. Influencing decision-making and implementation** | | |
| **Who do they collaborate with?** | Like-minded civil servants<br><br>Civil society | Institutional political actors:<br>• Political parties<br>• Civil servants<br><br>Social movements |
| **What kind of outcomes do they pursue?** | Tangible changes in concrete projects:<br>• Decision-making processes or regulations<br>• Official communications | Changes in citizens lived experiences:<br>• Participatory governance<br>• Policies |

### 4.1.1 Thinking and discourses

**Which audiences do public officials address?**

Civil servants and elected officials tend to focus on different audiences to raise awareness and support for post-growth approaches. Civil servants mainly aim to promote post-growth arguments by making them more widespread within their institution. For example, Interviewee 18 (CS/NL/N) reported that in their ministry they developed *"a 'serious game' [...] That's just to raise awareness that any project is much more complicated than you might imagine, 'cos you pull a string here and something happens on the right, which you don't know about, you don't understand the causal effects and relations"*. They also *"set up an innovation academy, which is an internal academy, so to speak, which does once-a-month sessions linked to innovation and post-growth as well"*.

In contrast, elected officials tend to address the broader public, external stakeholders, and other elected officials. Interviewee 33 (EO/ES/R) saw outward-facing communication as a priority *"because it is the public that are going to press, that are going to be the reason that the politicians will implement some things"*. Interviewee 32 (EO/ES/L) stated that *"Actually, the people with [whom] we don't have to talk [are] our people, because they already know [...] what we have to do and what we will do is start going to [...] entrepreneurs groups, groups of businessmen, groups or local commerce, [...] everyone that [...] they haven't heard maybe the word [degrowth]"*.

Elected officials also aim to influence other elected officials in parliamentary or local council debates. For instance, referring to the Beyond Growth conferences at the European Union level, Interviewee 34 (EO/ES/EU) described how they helped to bring post-growth ideas to the debates in the parliament: *"So they [parliamentarians] have to talk about degrowth, they have to talk about post-growth, and including if you don't agree [...] I think that's kind of a big success in that way. It doesn't mean the European Parliament is degrowth-oriented, not at all [...]. But the fight is here"*. Interviewee 33 (EO/ES/R) explained that to them the parliament "*is the most important place to introduce that kind of speech"*. Since parliamentary debates are often publicly available, they serve a double function of also addressing stakeholder groups and voters more broadly. Some elected officials said they also aim to influence the media narrative which was often perceived as hostile towards alternative economic approaches, *"But you have to make sure to explain it very simply there"* (Interviewee 25/EO/AT/N).

**Which terminology do public officials use?**

Civil servants and elected officials both adapt the terminology they use according to the context. Overall, the civil servants in our sample tended to avoid using the term "post-growth" (or degrowth). They fear that using "post-growth" would lead to their exclusion from debates and provoke strong opposition. Interviewee 2 (CS/AT/L) noted that they *"would speak very little about post-growth. Because that language is understood by a niche clientele"*. Interviewee 7 (CS/AT/N) did not *"think [that] it helps if you just say, okay, economic growth is evil and all that. Because then the other person just closes the door and doesn't want to talk to you anymore, because you're taking such an extreme point of view"*. Instead, they find the use of the term "wellbeing economy" particularly useful because *"the wellbeing economy is much easier to communicate [...]. That it is the much more apolitical term, which many* *people can identify with more, so to speak"*. In addition, multiple post-growth-minded interviewees said they use green-growth-aligned terminology because they feel that this is the more feasible approach.

In contrast, some elected officials said that they do directly criticize economic growth in their public appearances. Interviewee 37 (EO/UK/R) described that *"in the context of the debating chamber, in parliament, for instance, I'm not going to change anybody's mind in that space. But I still think it's important for me to say very, very overtly, 'Economic growth is failing'. [...] I think it's important that I say that because it holds that space open [...] for an alternative in our political discourse".* Interviewee 27 (EO/HR/L) would argue that *"we don't need the growth anymore, and the growth as such can be detrimental for the planet, for the economies and for the society as such".* However, they said that they would make these kinds of arguments *"not as a member of the city council, but more as a public sociologist [who] speaks about those kinds of topics".*

Other elected officials found that it is *"better not to talk about post-growth as such, but [...] speak about the practical changes in as much grounding to the practical reality as possible"* (Interviewee 30/EO/FI/L). Thus, they aim to connect their arguments to the daily life and experiences of citizens to make their demands more relatable. For instance, Interviewee 29 (EO/HR/N) said that *"when you talk about big topics, just people can't relate to that and it's more of an academic discussion than it is something that the people actually see as relevant. They see it as relevant when you talk to them about transport, about social aid, about, I don't know, childcare [...]".*

**What kind of information do public officials perceive as most legitimate?**

Many civil servants stated that they find technical evidence from governance tools, indicator frameworks, and in-house studies useful to back up their position because *"we need some tools that [...] are based on science, so nobody can complain about that"* (Interview 20/CS/ES/R). In addition, Interviewee 3 (CS/AT/L) found that *"it's important to be able to reference some paper that comes from policymakers. And then we can say that it's our job to implement this"*.

Some elected officials found that scientific evidence on the urgency of multiple crises was useful to make the case for post-growth: *"The only times I've like managed to [...] ... change some little things, it's with that kind of scientific background that helps [...] them understand some things"* (Interview 33/EO/ES/R). In addition, elected officials referred to best practice examples of the changes that they were aiming to promote to justify that they were possible and advantageous: *"For example, on the reduction of working hours, we have a lot of [...] companies working on that, and that's one of the best added value that we have right now, because when the company says in a world so dominated by economy, that it works, well, a lot of people in the society, most of all the politics, they can't say, [...] 'you can't do that, that's impossible,' no, because it works"* (Interview 34/EO/ES/EU). Moreover, Interviewee 26 (EO/HR/L) stated that if post-growth was in an elected official's political "*programme, this is kind of an ultimate argument to say, 'Okay, we are elected for this, and this is what should be done'.*"

### 4.1.2 Decision-making and implementation

**Who do public officials collaborate with?**

We find that civil servants tend to promote post-growth approaches by building a community with like-minded civil servants to support each other. For instance, Interviewee 10 (CS/HR/L) said: *"you see how other people think about some subjects and when you see they're thinking the same as you, you connect with them and collaborate. I don't know, when you propose something, you ask them, they say, 'Okay, I think also that we should change it,' and then you come to do a meeting and there is two of us. You're not alone".*

Several civil servants also described actively inviting civil society actors to events to bring their more transformative positions to the discussion, allocate funding to Non-Governmental Organisations (NGOs), or provide them with information. For example, Interviewee 4 (CS/AT/N) recalled that *"the civil society actors from the environmental NGOs were very important partners. These organisations are able to formulate very clear statements without getting pushback from the top. [...] At the same time, we are in control of the budgets, and the NGOs are thankful for every cent. [...] We also put together an in-house commission that included all department heads of the [ministry]. [...] And from the start, the commission had two representatives from environmental umbrella associations, that had many NGOs as members".*

Elected officials collaborate with a variety of institutionalized political actors *"because it is never that the civil servants are only pushing for something, or the politicians are pushing for something, it is a very weird combination of things, and I think governance is more often than not much more unstructured and random than we think it is. It is a lot about personal relationships"* (Interview 30/EO/FI/L). Interviewee 34 (EO/ES/EU) agreed that it is important *"to create alliances outside of your bubble, [...] what we did was to look for other MEPs [Members of the European Parliament], for example, where...I don't say who were absolutely 100% post-growth thinkers, but at least who were able or wanted to have this debate".*

Like civil servants, many elected officials collaborate with social movements in formal and informal ways. For example, Interviewee 29 (EO/HR/N) reported that "*if they have different policies that they want us to allocate in the Croatian parliament, we meet with them. Also on the city level, we meet with them. Sometimes they are part of our working groups.*"

**What kind of outcomes do public officials pursue?**

Regarding the outcomes that they pursue, civil servants tend to aim for specific procedural changes in decision-making processes or regulations. For example, governance tools and alternative measurement metrics are central instruments for civil servants to foster post-growth approaches. Interviewee 16 (CS/NL/L) explained that *"through our budgeting process, when the different departments make their budgets, they now have to indicate what they are going to do on sustainability [...] I see this really as a systemic change in how the municipality works"*. Moreover, they introduce and adapt the terminology in communications. Interviewee 7 (CS/AT/N) gave the example of the reports of the Intergovernmental Platform on Biodiversity and Ecosystem Services and the Intergovernmental Panel on Climate Change where *"there are the so-called 'indirect drivers'. These are what affects economic growth and so on and so forth. I find it problematic that they are called 'indirect drivers' as opposed to 'direct drivers', so, rather the root causes or just the underlying problems. I think it would be very important to make this connection in this environmental assessment report"*.

Instead of specific tools, most elected officials aim to foster participatory governance approaches and the implementation of alternative policies to change decision-making and ensure public support. Some elected officials discussed participatory governance mechanisms as a tool to support public and political acceptance of post-growth policies. For example, Interviewee 40 (EO/UK/N) suggested that *"people have lost the sense of political agency, and possibly the most important thing we can do is restore people's sense of political agency [...] I don't believe [in] giving up on representative democracy, but I think we also need a layer of participatory democracy, [...] those constitutional conventions, those sort of things"*. Interviewee 38 (EO/UK/R) outlined that *"one of the reasons that this [the Future Generations' Act] is attractive or palatable in the context of the people of Wales is the people of Wales have been involved in discussions about what they want their future to look like on a number of occasions [...]."*

Similarly, some elected officials stated that they aim to implement policies which would let citizens directly experience the advantages of post-growth policies in their daily life which then potentially makes them more supportive of post-growth positions in the future. For example, Interviewee 28 (EO/HR/N) stated that *"it's not about the theory or the narrative or arguments, it's about people feeling that after those policies are introduced, [...] their lives are better."*

## 4.2 Capacity of the state

Public officials' cautious and incremental strategies reflect the levers but also the overall limited capacity of the state to support post-growth approaches under the current growth paradigm. First, public officials' reported the dominance of growth-focused ideology: *"It's presented as completely logical that the economy has to grow. That's not even questioned."* (Interview 3/CS/AT/L), and "*unfortunately, it has to be said that the growth-driven ideology usually wins*" (Interview 7/CS/AT/N). Accordingly, some public officials perceive the power of their institution as limited in contrast to economic actors: *"I think there is still a huge, yeah, a huge gap between power of the economic world and what we can do as a public service"* (Interview 15/CS/FR/L). Second, public officials experience monetary growth-dependencies of the state: *"Our budget rises and falls in accordance with economic growth"* (Interview 2/CS/AT/L). For example, Interviewee 26 (EO/HR/L) reported that *"the way the financial construction of short-term but also mid-term investments in infrastructure are planned, we need to continue this kind of indebtedness and [...] these decisions to continue these kind of financing basically implied also in a way economic growth".*

Cooperation with civil society actors was a way for public officials to challenge the ideological constraints in some circumstances since civil society actors are not bound to the same institutional limits: *"it's very relevant that civil society always has a broader imagination than politics [...] if you don't have this kind of imagination and criticism of how actual day-to-day politics works, then you're going to at best have the status quo of social change. [...] I think that in a way non-institutional actors always have to be more ambitious than are actually the limits of political action"* (Interviewee 29/EO/HR/N). On the one hand, public officials collaborated with civil society actors in institutionalized ways. For example, Interviewee 8 (CS/AT/N) described proactively initiating post-growth-leaning projects with a civil society organization: *"We thought it was important, so we approached them and developed a project, and provided financial support."* In addition, by inviting civil actors to internal events, civil society actors promoted post-growth discourses and proposals directly inside the institutions (Section 4.1.2). On the other hand, some public officials focus on outward facing communication (Section 4.1.1) and collaborate with outside actors to increase the external pressure on their institution (Section 4.1.2). For example, Interviewee 26 (EO/HR/L) described that they collaborate with *"people who can contribute to putting a certain topic in the mainstream and to make it visible, to kind of create also public pressure on certain topics. So when some decisions are not all entirely going to be made in the political terms because of the pragmatics, we also try to influence the public sphere in order to address this issue so it becomes a pressure also from different aspects."*

Thus, the strategies point to the importance of empowering post-growth proponents in civil society in order to increase the pressure on and thereby the capacity of public officials and government institutions to advance post-growth approaches: "*we need braver decision makers, we need much more community involvement in decision making, we need much more support and capacity building at a community level. And we need a complete rebalancing of power, both financial clout, but also the sort of soft power around access to decision making"* (Interviewee 37/EO/UK/R). One way to increase the capacity to advance post-growth approaches could be by deepening democratic participation and governance mechanisms (Section 4.1.2) *"because [then] the government will always have, like, an* *excuse, right, 'The citizens are saying this, so I have to implement this. It's not, like, my idea, they are the citizens who are deciding' [...] So yeah, I think that...not only assemblies, but also referendums and consultations are key"* (Interviewee 19/CS/ES/R).

## 5. Discussion

Overall, we find that public officials aim to influence thinking and discourses to foster broader support for post-growth approaches, while also trying to impact concrete decision-making and policy implementation. However, we find several key differences in strategies between civil servants and elected officials. With regards to how open public officials chose to be in their support for post-growth we find that civil servants tend to opt for more indirect and subtle strategies that avoid directly criticising economic growth and using a post-growth framing. Elected officials are more outspokenly critical of economic growth. Their main audiences also tend to differ: civil servants are more likely to focus on audiences within their institutions while elected officials tend to target other elected officials and the broader public. In relation to influencing decision-making and implementation, civil servants lean towards building collaborations with their colleagues and with civil society organizations to influence decision-making inside their institutions, whereas elected officials tend to collaborate with a wider breath of political actors and social movements. Civil servants are more interested in the implementation of governance tools and alternative measurement metrics whereas elected officials tend to aim for expanding the possibilities for participation in democratic processes and the implementation of new policies.

Our analysis has several limitations that should be acknowledged. First, the research design of a cross-sectional interview study allows only for limited trust-building between interviewee and interviewer. Thus, it can be challenging to gain transparent information by public officials about all strategies that they employ. Some of the study participants might therefore not have told us about some more confrontative or subversive strategies that they have pursued. Second, the snowball sampling led to us interview multiple individuals from some but not all institutions, giving us varying levels of detail across institutions. Third, our sample only includes elected officials who perceive a post-growth position in line with their political mandate. Arguably there might be post-growth-minded elected officials who do not perceive that position to be in line with the public position they were elected on. Thus, they might pursue more indirect strategies similar to those that civil servants in our sample tended to apply. Fourth, due to our focus on commonalities across a wide range of contexts we lack in-depth analysis of the country, political and institutional settings, as well as situational knowledge about the various actions described by change agents. Lastly, it is out of the scope of the study to provide a differentiation of the actions employed depending on the seniority and job responsibility of interviewees.

Notwithstanding these limitations, our study makes three important contributions to the literature on the role of the state in post-growth transformations: First, we provide empirical insights on the strategies of public officials to foster post-growth approaches from across Europe and on different governance scales. Second, we showcase the differences in strategies between civil servants and elected officials. Third, we argue that while these strategies demonstrate that there are levers for state actors to promote post-growth approaches, our findings also underscore the limited capacity of the state to advance post-growth approaches under the current growth paradigm. The strategies showcase the importance of cooperation with civil society actors to challenge institutional limitations and advance symbiotic and interstitial strategies simultaneously. The findings can provide starting points on how to leverage the role of public officials to foster post-growth from within government institutions.

For example, the strategy of coalition building that our interviewees described addresses two barriers to post-growth approaches (Angresius et al., 2025; Hayden, 2025; Mason & Büchs, 2023; Schmid, 2025): First, building alliances within government institutions aims to strengthen post-growth positions within the institutions and to overcome siloed ways of working. Second, coalition building with civil society actors is a way to challenge the institutional structures that public officials are embedded in. On the one hand, inviting civil society actors to meetings and events allows them to bring critical perspectives that they could not voice themselves. On the other hand, as described by Wright (2010), interstitial actors also rely on support by and cooperation with actors in established political institutions to sustain their initiatives. Therefore, coalition building beyond government institutions mutually strengthens symbiotic and interstitial actors which enhances the capacity of the state to advance post-growth approaches. As outlined in the state-optimistic post-growth literature societal (D'Alisa & Kallis, 2020a; Koch, 2022), our findings underscore the importance of close links between civil society and state actors to foster large-scale.

While we investigated individuals' strategies, Wright (2019) points out that only collective agency would be able to challenge powerful institutionalized interests and established paradigms. Our findings suggest that the level of openness to criticize economic growth relates to the level of collective agency that public officials experience. Elected officials in our study were more likely than civil servants to openly adopt positions that are perceived as "radical". This was the case regardless of whether they were in government or in opposition, and across governance scales. Our analysis suggests that one characteristic that elected officials who openly criticized economic growth had in common was that they felt that these positions reflected their political program and mandate. Thus, we suggest that beyond the formal legitimacy of their position, speaking for their party position equips the individual elected officials with a sense of collective agency which allows them to challenge the economic growth paradigm more openly.

In contrast, while many civil servants were looking to build (informal) networks and connections of support with like-minded actors, in our interviews their reference points for legitimizing their post-growth stance were technical information and established policies and legislation. As showcased by the example of the letter in Vestergaard and Schmid (2026), joining together as a group with shared values and interests could allow post-growth-minded civil servants to be more openly confrontative in their actions. Therefore, focusing on extending the (informal) networks within their institutions could be a promising avenue for civil servants to extend their sense of collective agency.

One barrier to creating collective agency among civil servants could be the dominant self-understanding as apolitical actors in the given institutional structures. In line with the inner barriers that Lamker and Dieckhoff (2022) identified for post-growth planners, some civil servants in our study tried to be perceived as neutral and described attempts to depoliticize their actions, e.g. by referring to scientific evidence and established agreements. This self-understanding could be interpreted as a barrier to civil servants conducting more confrontative mobilizations, and politicizing decision-making (Akba, 2023; Feola, 2025). In addition, the tendency of civil servants to promote post-growth approaches more indirectly could make some post-growth government initiatives more prone to co-optation by growth-focused actors and interests (Hayden, 2025; Schmid, 2025). However, we would like to point out that many civil servants in our study described opting for the more subtle actions because they perceived this avenue to be either more effective, or even the only feasible one. Our study does not allow us to draw conclusions on the (relative) effectiveness of either more confrontative or more subtle approaches. Rather, it underscores the importance of engaging in dialogue with and harnessing the practical knowledge of public officials on avenues for next-best policy changes in their specific contexts.

Wright (2010) argues for strengthening the democratic capacity of citizens as a means and as an end for symbiotic transformations across contexts.[2] Some elected officials in our interviews described respective ambitions to boost citizens' influence in formal policymaking spaces. There are multiple proposals in the post-growth literature for how public participation could be formalized, for example through deliberative citizen forums (Koch, 2020a) and a dual strategy of policy formulation, combining expert knowledge with the practical knowledge of citizens (Gough, 2017). Furthermore, focusing on the implementation of specific services and policies to provide citizens with lived experiences of post-growth approaches could be a promising avenue to increase political support (Bärnthaler, 2024). In the interviews, working-time reduction was mentioned as one example of a potentially appealing post-growth policy which would furthermore have the capacity to free up time for democratic participation (Fitzpatrick et al., 2022; Oberholzer, 2023).

[2] The literature on strategies for post-growth transformations discusses similar proposals as non-reformist reforms (Kallis et al., 2020). Applied to post-growth, these are *"radical reforms that transform existing institutions by creating growth independence and improving sustainability and justice"* (Feola, 2025, p. 2). Non-reformist reforms contribute to changing essential characteristics of the status quo, which eventually accumulate in transformative institutional change (Akba, 2023; Gorz, 1968).

Future research might explore the effectiveness and impact of strategies to foster post-growth approaches in different contexts through case studies. In addition, future studies might examine which terminology and framing are most useful for public officials to communicate post-growth approaches within and beyond political institutions. In addition, future survey studies could quantitatively compare the prevalence of post-growth mindsets and strategies among public officials across country context and governance scales.

## 6. Conclusion

In this study, we have investigated three research questions: (1) How do post-growth-minded public officials promote post-growth approaches in their work? (2) How do strategies differ between civil servants and elected officials? (3) What do these strategies reveal about the capacity of the state to advance post-growth transformations? To answer these questions, we analysed 41 interviews with civil servants and elected officials in seven European countries from local to European governance scales.

We find that civil servants and elected officials aim to influence the thinking and discourses around post-growth, as well as specific decision-making and the implementation of policies. Cooperating with civil society actors was a way for public officials to navigate the tension of challenging institutional structures while being embedded in them. The strategies diverged between civil servants and elected officials in line with their institutional roles. Civil servants primarily rely on formal instruments, including governance tools and the organization of events, and justify their positions by referring to legislation and formal agreements. In contrast, elected officials are more open to criticizing economic growth, with a focus on enhancing democratic participation and implementing policies.

While our findings demonstrate some of the levers that state actors can draw on to foster change, they also underscore their limited capacity to advance post-growth approaches under the current growth paradigm. Thus, we argue that coalition building with civil society actors needs to be front and center for symbiotic and interstitial strategies to be mutually enhancing. In addition, we suggest that the level of openness to criticize economic growth relates to the level of collective agency that public officials experience. Therefore, to challenge growth-focused ideologies and approaches in government institutions, strengthening the collective agency of post-growth minded public officials could be key.

## Acknowledgments

We would like to thank all interviewees who have given their time generously to participate in this study. In addition, we would like to thank all individuals who supported us in identifying potential interviewees, in particular Ernest Aigner. Ethics approval was obtained from the University of Barcelona (Bioethics Commission), and from the University of Leeds (Faculties of Business, Environment, and Social Sciences Ethics Committee, reference 0937), covering full informed consent from interview participants and anonymization of interview data in research outputs.

This research received funding from the European Union's Horizon Europe research and innovation programme under grant agreement number 101094211 (ToBe: "Towards a Sustainable Wellbeing Economy: Integrated Policies and Transformative Indicators") and UK Research and Innovation (UKRI) under grant agreement number 10052343. Views and opinions expressed are, however, those of the authors only and do not necessarily reflect those of the European Union or UKRI. Neither the European Union, the granting authority, nor UKRI can be held responsible for them.

## Data availability statement

The data set underlying this study is not yet publicly available but will be published in a data repository in the future.

## Author contributions

**Laura Angresius:** Conceptualization, Methodology, Investigation (data collection), Formal analysis, Writing – Original Draft, Writing – Review & Editing; **Milena Büchs:** Conceptualization, Methodology, Investigation (data collection), Writing – Review & Editing, Supervision, Funding acquisition; **Daniel W. O'Neill:** Writing – Review & Editing, Supervision, Funding acquisition

## Disclosure statement

The authors report there are no competing interests to declare.

## Annex

**Table 1A:** Overview of interviews

| # | Position | Country | Governance scale | Interview code[†] |
|---|---|---|---|---|
| **1** | Civil servant | Austria | Local | 1/CS/AT/L |
| **2** | Civil servant | Austria | Local | 2/CS/AT/L |
| **3** | Civil servant | Austria | Local | 3/CS/AT/L |
| **4** | Civil servant | Austria | National | 4/CS/AT/N |
| **5** | Civil servant | Austria | National | 5/CS/AT/N |
| **6** | Civil servant | Austria | National | 6/CS/AT/N |
| **7** | Civil servant | Austria | National | 7/CS/AT/N |
| **8** | Civil servant | Austria | National | 8/CS/AT/N |
| **9** | Civil servant | Austria | National | 9/CS/AT/N |
| **10** | Civil servant | Croatia | Local | 10/CS/HR/L |
| **11** | Civil servant | Finland | National | 11/CS/FI/N |
| **12** | Civil servant | Finland | National | 12/CS/FI/N |
| **13** | Civil servant | Finland | National | 13/CS/FI/N |
| **14** | Civil servant | Finland | National | 14/CS/FI/N |
| **15** | Civil servant | France | Local | 15/CS/FR/L |
| **16** | Civil servant | Netherlands | Local | 16/CS/NL/L |
| **17** | Civil servant | Netherlands | Local | 17/CS/NL/L |
| **18** | Civil servant | Netherlands | National | 18/CS/NL/N |
| **19** | Civil servant | Spain | Regional | 19/CS/ES/R |
| **20** | Civil servant | Spain | Regional | 20/CS/ES/R |
| **21** | Civil servant | Spain | National | 21/CS/ES/N |
| **22** | Civil servant | United Kingdom | Regional | 22/CS/UK/R |
| **23** | Civil servant | United Kingdom | Regional | 23/CS/UK/R |
| **24** | Civil servant | United Kingdom | Regional | 24/CS/UK/R |
| **25** | Elected official | Austria | National | 25/EO/AT/N |
| **26** | Elected official | Croatia | Local | 26/EO/HR/L |
| **27** | Elected official | Croatia | Local | 27/EO/HR/L |
| **28** | Elected official | Croatia | National | 28/EO//HR/N |
| **29** | Elected official | Croatia | National | 29/EO/HR/N |
| **30** | Elected official | Finland | Local | 30/EO/FI/L |
| **31** | Elected official | Finland | Local | 31/EO/FI/L |
| **32** | Elected official | Spain | Local | 32/EO/ES/L |
| **33** | Elected official | Spain | Regional | 33/EO/ES/R |
| **34** | Elected official | Spain | European Union | 34/EO/ES/EU |
| **35** | Elected official | United Kingdom | Local | 35/EO/UK/L |
| **36** | Elected official | United Kingdom | Local | 36/EO/UK/L |
| **37** | Elected official | United Kingdom | Regional | 37/EO/UK/R |
| **38** | Elected official | United Kingdom | Regional | 38/EO/UK/R |
| **39** | Elected official | United Kingdom | Regional | 39/EO/UK/R |
| **40** | Elected official | United Kingdom | National | 40/EO/UK/N |

| **41** | Elected official | United Kingdom | European Union | 41/EO/UK/EU |
|---|---|---|---|---|

† The interview code is made up of the following aspects: **(1) Interview number. (2) Position** CS: Civil servant, EO: Elected official. **(3) Country:** AT: Austria, HR: Croatia, FR: France, FI: Finland, NL: Netherlands, ES: Spain, UK: United Kingdom **(4) Governance level:** EU: European Union, N: National, R: Regional, L: Local.